# Understanding kinetic interactions between $NO_x$ and $C_2$-$C_5$ alkanes and alkenes: The rate rules and influences of H-atom abstractions by $NO_2$

Hongqing WU[a,^], Ruoyue TANG[a,^], Xinrui REN[a], Mingrui WANG[a], Guojie LIANG[a], Haolong LI [a], Song CHENG[a,b]∗

[a] *Department of Mechanical Engineering, The Hong Kong Polytechnic University, Kowloon, Hong Kong SAR, China*
[b] *Research Institute for Smart Energy, The Hong Kong Polytechnic University, Kowloon, Hong Kong SAR, China*

^ Both authors contribute equally to this paper.
∗ Corresponding authors: Song Cheng
Phone: +852 2766 6668
Email:songcheng@polyu.edu.hk

**Novelty and Significance Statement**

This study presents a comprehensive study on the H-atom abstraction by $NO_2$ from $C_2$-$C_5$ alkanes and alkenes, encompassing 15 hydrocarbons and 3 $HNO_2$ isomers with 45 reactions. Through a combination of high-level quantum chemistry computation, rate extraction, kinetic analysis and model implementations, the rate rules for this type of reactions from alkanes and alkenes are determined. These rate rules can be used to reliably derive rate parameters over a wider range of hydrocarbons. The chemical kinetic modeling and the in-depth analysis also reveal the strong influences of this type of reaction on $NO_X$/hydrocarbon interactions and model reactivity, and emphasize the need to accurately describe their kinetics in existing models, which can now be sufficiently achieved through the results presented in this study.

**Author Contributions**

H.W. R.T. and S.C. conceived and designed the framework and performed the calculations, with technical inputs from X.R., M.W., G.L. and H.L. All authors contributed to writing the paper.

## Abstract:

This study aims to reveal the important role and the respective rate rules of H-atom abstractions by $NO_2$ for better understanding $NO_X$/hydrocarbon interactions. To this end, H-atom abstractions from $C_2$-$C_5$ alkanes and alkenes (15 species) by $NO_2$, leading to the formation of three $HNO_2$ isomers (TRANS_HONO, $HNO_2$, and CIS_HONO) and their respective products (45 reactions), are first characterized through high-level quantum chemistry computation, where electronic structures, single point energies, C-H bond dissociation energies and 1-D hindered rotor potentials are determined at DLPNO-CCSD(T)/cc-pVDZ//M06–2X/6−311++g(d,p). The rate coefficients for all studied reactions, over a temperature range from 298.15 to 2000 K, are computed using Transition State Theory with the Master Equation System Solver program. Comprehensive analysis of branching ratios elucidates the diversity and similarities between different species, $HNO_2$ isomers, and abstraction site, from which accurate rate rules are determined. Incorporating the updated rate parameters into a detailed chemical kinetic model reveals the significant influences of this type of reaction on model prediction results, where the simulated ignition delay times are either prolonged or reduced, depending on the original rate parameters presented in the selected model. Sensitivity and flux analysis further highlight the critical role of this type of reaction in affecting system reactivity and reaction pathways, emphasizing the need for adequately representing these kinetics in existing chemistry models. This can now be sufficiently achieved for alkanes and alkenes through the results from this study.

# 1. Introduction:

Nitrogen oxides ($NO_x$), including $NO$, $NO_2$, and $N_2O$, are critical pollutants regulated for their adverse environmental impacts. They can be formed during the combustion processes of ammonia, hydrogen and typical hydrocarbons in air [1]. To mitigate $NO_x$ formation, exhaust gas recirculation (EGR) is employed to reduce the combustion temperature so as to reduce thermal $NO_x$ formation, where part of the exhaust $NO_x$ is introduced into the combustion chamber to mix with a fresh charge. Although the $NO_x$ are the minor species in the mixture with very low concentrations, they can significantly alter the original reaction pathways, leading to changes in system reactivity [2].

In recent years, the interactions between $NO_x$ species and typical fuel components have been increasingly studied. To date, experimental studies of interaction chemistry between $NO_x$ species with typical hydrocarbons have already been reported in jet-stirred reactors (e.g., methane, ethane) [3-5], laminar flow reactors (e.g., methane, ethane, methanol) [6-8], rapid compression machines (e.g., methane, ethane, ethylene, propene, isobutene, RD5-87 and PACE-20) [2,9,10], and a circular flat flame burner (acetylene) [11]. These studies generally emphasized the role of $NO_x$ in promoting fuel reactivity at low concentrations while inhibiting fuel reactivity at high concentrations. The underlying mechanism of $NO_x$ interactions has been further elucidated through sensitivity and flux analysis, among which the primary key reactions are revealed as:

$$NO + H\dot{O}_2 \leftrightarrow NO_2 + \dot{O}H \qquad (R1)$$

$$\dot{R} + NO_2 \leftrightarrow R\dot{O} + NO \qquad (R2)$$

$$\dot{H} + NO_2 \leftrightarrow NO + \dot{O}H \qquad (R3)$$

$$RH + NO_2 \leftrightarrow \dot{R} + TRANS\_HONO/CIS\_HONO/HNO_2 \qquad (R4)$$

Despite that comprehensive studies on reactions R1, R2 and R3 have been conducted, the kinetics and rate rules of R4 have been less studied, leaving the influences from $NO_2$ interactions underexplored. It is obviously necessary to understand the interaction kinetics of $NO_2$, as NO is found to quickly convert to $NO_2$, even at room temperature [12]. A lack of such understanding casts uncertainties on understanding the $NO_x$ addition effects, which might further result in ill-conditioned modeling frameworks. $NO_2$ contains an unpaired electron, making it prone to H-atom abstraction reactions [12]. This type of reaction has already been identified in past studies that studied $NO_x$ blending effects. For instance, Cheng et al. [2] revealed that the H-atom abstraction from iso-butene, i.e., $IC_4H_8 + NO_2 \leftrightarrow IC_4H_7 + HONO$, greatly promote the overall autoignition reactivity of iso-butene using experiments from a twin-piston rapid compression machine. Deng et al. [13] investigated the $NO_2$ addition effects on autoignition behavior of propylene, and the importance of H-atom abstraction from propylene at high $NO_2$ addition levels was highlighted. Zhang et al. [14] found that $NO_2$-driven H-atom abstraction reactions have a promoting effect on methane and ethane by measuring the ignition delay times of $CH_4/NO_2/O_2/Ar$, $C_2H_6/NO_2/O_2/Ar$, and $CH_4/C_2H_6/NO_2/O_2/Ar$ mixtures in a shock tube. Rasmussen et al. [15] reported the importance of H-atom abstraction reactions between $CH_3O/CH_2O$ and $NO_2$ at low temperatures. These studies confirmed the significant contributions from H-atom abstractions from fuel molecules for a diversified range of fuels including alkenes, alkanes and aldehydes, highlighting the importance of this type of reaction.

There are a few theoretical studies on H-atom abstraction reactions by $NO_2$. Chai et al. [16] calculated the rate coefficients for H-atom abstraction from several alkanes and alkenes by $NO_2$ at the CCSD(T)-F12a/cc-pVTZ-f12//B2PLYPD3/cc-pVTZ level of theory. Wang et al. [17] calculated the H-atom abstraction from n-decane by $NO_2$ by CBS-QB3//M06–2X/6–311++G (d,p) method. Nevertheless, there remains a lack of comprehensive and systematic theoretical studies on H-atom abstraction reactions by $NO_2$, particularly for different H-atom sites on different molecules and with considerations of complexes. As such, the rate rules for such reaction from different hydrocarbons remain unclear. Additionally, previous studies have not fully explored the impact of these reactions on model predictions nor adequately considered the effects of $HNO_2$ isomers within reaction systems.

Therefore, this study aims to address these gaps by: (a) conducting a detailed theoretical investigation of H-atom abstractions from $C_2$-$C_5$ alkanes and alkenes by $NO_2$ that form $HNO_2$, TRANS_HONO and CIS_HONO, with the consideration of complexes whenever applicable; (b) reveal the branching ratios of the three pathways forming the three $HNO_2$ isomers for the selected species at different H-atom sites and different size of molecules; (c) determine the rate rules for this type of reactions for alkanes and alkenes; and (d) systematically analyze the effects of these reactions on model prediction results.

## 2. Computational methods:

Supported by recent study validations [18-21], the M06-2X method [22] coupled with the 6-311++G(d,p) basis set [23-25] is utilized for optimizing geometric structures, calculating vibrational frequencies, zero-point energies (ZPE), and scanning dihedral angles for all species, complexes, and transition states (TS). Leveraging this high-level method, the structures of transition states are determined through scanning the C-H bond lengths via relaxed scans. The transition states are confirmed with an imaginary frequency, as well as using intrinsic reaction coordinate (IRC) to ensure that these transition states accurately represent the pathways linking the respective reactants and products. Additionally, the dihedral angles are rotated in 18 incremental steps of 20 degrees each step to determine the hindered rotor potentials. According to Zhao and Truhlar [22], scaling factors of 0.983 for harmonic frequencies and 0.9698 for ZPEs are suitable under the M06-2X method, which are adopted in the study. The DLPNO-CCSD(T) functional [26,27] with the cc-pVDZ basis set is employed to obtain single-point energies (SPEs) and T1 diagnostics. The T1 diagnostics for all species are below 0.028, affirming the correctness of molecular structure and the suitability of the M06-2X method for these reaction systems.

For the calculation of rate constants, Transition State Theory (TST) is utilized, implemented through the Master Equation System Solver (MESS) program [28]. This analysis incorporates a one-dimensional (1-D) Eckart model [29], tailored to calculate the rate coefficients for hydrogen abstraction reactions. Additionally, hindered rotor potentials obtained from one-dimensional scans are used to adjust the frequencies of lower-frequency modes, ensuring more accurate rate constant calculations. The temperature range is set from 298.15 to 2000 K. All rate constants are fitted to the modified Arrhenius equation:

$$k = AT^n \exp(-E_a/RT)$$

where $A$ is the frequency factor, $T$ is the temperature, $n$ is the temperature-dependent curvature factor, $E_a$ is the activation energy, and $R$ is the ideal gas constant.

To reveal the influence caused by these newly calculated reactions on autoignition characteristics of investigated species, IDT is calculated using closed homogenous batch reactor module in Zero-RK [30] using a detailed chemistry model developed by Cheng et al. [31].

## 3. Results and discussion

3.1. Species and reaction sites

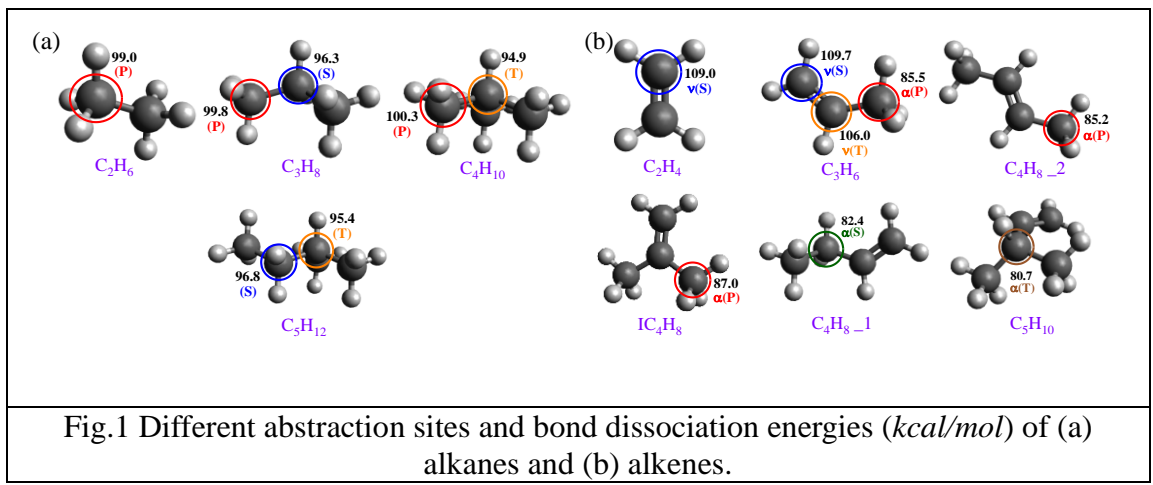

Fig.1 Different abstraction sites and bond dissociation energies (*kcal/mol*) of (a) alkanes and (b) alkenes.

This study investigates the reactions of four alkanes and five alkenes with $NO_2$, leading to the formation of three types of $HNO_2$ isomers (TRANS_HONO, $HNO_2$, and CIS_HONO) and their corresponding products, encompassing a total of 45 reactions. The reaction sites for different reactants and the associated bond dissociation energies are depicted in Figure 1. Concerning the C-H bond sites within the molecules, these are categorized based on the carbon atom position as primary (labeled P), secondary (labeled S), and tertiary (labeled T). As for C=C functional group, the carbon atoms on the C=C double bonds are designated as ν, and those adjacent to C=C double bonds are identified as α.

Though reaction $RH \rightarrow \dot{R} + \dot{H}$, the bond dissociation energies (BDEs) of the C-H bonds in the reactants at 298K is calculated as:

$$BDE_{298}(R-H) = \Delta H^0_{f,298}(\dot{R},g) + \Delta H^0_{f,298}(\dot{H},g) - \Delta H^0_{f,298}(RH,g)$$

In alkanes, as depicted in Fig. 1(a), the differences in bond dissociation energies (BDEs) among them are minor, ranging from 1 to 5 kcal/mol. It is evident that the BDEs gradually decrease from the primary hydrogen atom (approximately 100 kcal/mol) to the secondary hydrogen atom (approximately 96 kcal/mol), with the tertiary hydrogen atom exhibiting the lowest BDE (approximately 95 kcal/mol). This trend underscores the influence of the hydrogen atom's position within the molecular structure on its bond strength. In the case of alkenes, as depicted in Fig. 1(b), the BDEs at the primary sites range from 85.2 to 87.0 kcal/mol, while the secondary and tertiary site is 82.4 kcal/mol and 80.7 kcal/mol, respectively. These values are lower than those of alkanes at corresponding sites, with a difference of about 14 kcal/mol, indicating that the C=C functional group markedly decreases the bond dissociation energy of hydrogen atoms on adjacent carbon atoms. For $C_3H_6$, the BDE at the ν(S) site is the highest, at approximately 109.0 kcal/mol, followed by ν(T) at 106.0 kcal/mol and α(P) at 85.5 kcal/mol, indicating that H-atom abstraction from α sites is facilitated compared to ν sites.

### 3.2. Potential energy surface

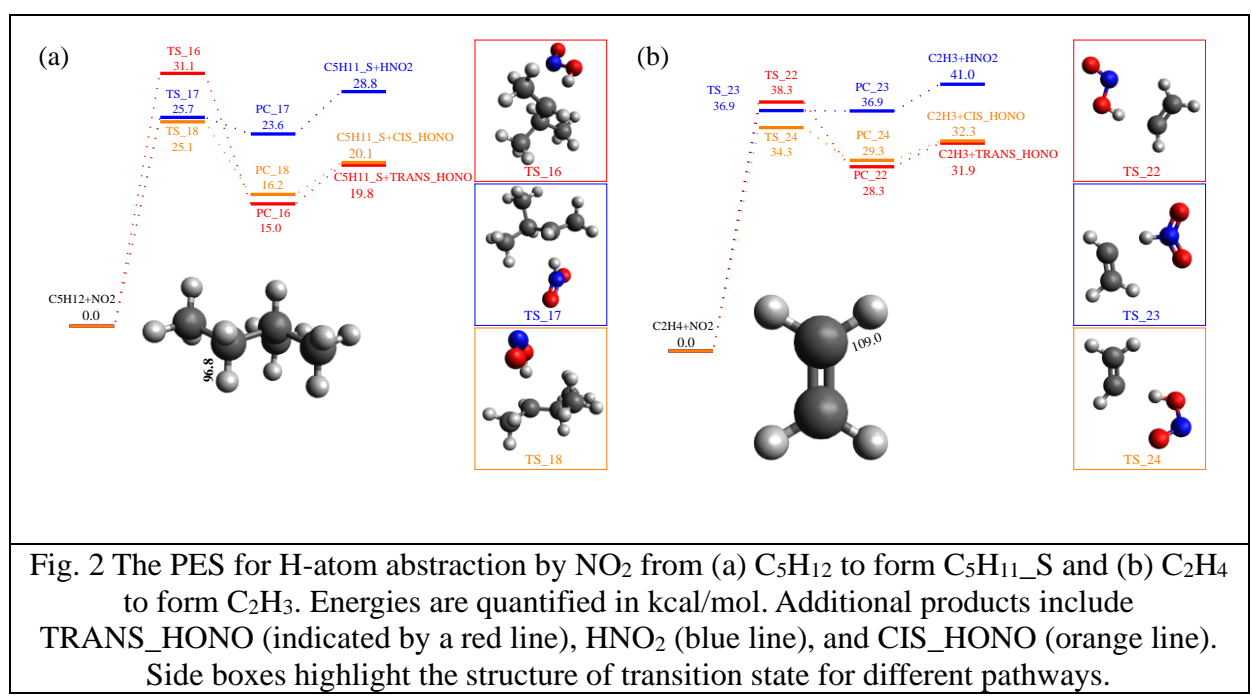

Fig. 2 The PES for H-atom abstraction by $NO_2$ from (a) $C_5H_{12}$ to form $C_5H_{11}\_S$ and (b) $C_2H_4$ to form $C_2H_3$. Energies are quantified in kcal/mol. Additional products include TRANS_HONO (indicated by a red line), $HNO_2$ (blue line), and CIS_HONO (orange line). Side boxes highlight the structure of transition state for different pathways.

The Potential energy surface (PES) for H-atom abstraction by $NO_2$ from $C_5H_{12}$ and $C_2H_4$ to form $HNO_2$ isomers and the relative product can be seen in Fig. 2(a) and 2(b). It is clear that the energy barriers of the reactions producing TRANS_HONO are consistently higher than those producing $HNO_2$ and CIS_HONO. Chai et al. [17] attribute this phenomenon to the reduced distance between RH and $NO_2$ in the formation of R and TRANS_HONO, compared to other reactions. This proximity increases Coulombic repulsion, consequently elevating the energy barrier. Additionally, the PES for all other species involved in this study is provided in the Supplementary Material (Figs. S1-S2), along with the T1 diagnostic values and optimized structures for all species.

Table 1. The relative energy for H-atom abstraction by $NO_2$ from different sites of alkanes and alkenes to form the respective products and $HNO_2$ isomers (TRANS_HONO, $HNO_2$, CIS_HONO). All values are in kcal/mol.

| No. | Reaction | Reactant | Transition state | Product complex | Product |
|---|---|---|---|---|---|
| Alkanes + $NO_2$ | | | | | |
| R1 | $C_2H_6+NO_2 \rightarrow C_2H_5\_P+TRANS\_HONO$ | 0 | 34.1 | 18.6 | 22.0 |
| R2 | $C_2H_6+NO_2 \rightarrow C_2H_5\_P+HNO_2$ | 0 | 29.8 | 27.2 | 31.0 |
| R3 | $C_2H_6+NO_2 \rightarrow C_2H_5\_P+CIS\_HONO$ | 0 | 28.2 | 19.3 | 22.3 |
| R4 | $C_3H_8+NO_2 \rightarrow C_3H_7\_P+TRANS\_HONO$ | 0 | 34.0 | 19.5 | 22.8 |
| R5 | $C_3H_8+NO_2 \rightarrow C_3H_7\_P+HNO_2$ | 0 | 29.9 | 27.9 | 31.8 |
| R6 | $C_3H_8+NO_2 \rightarrow C_3H_7\_P+CIS\_HONO$ | 0 | 28.0 | 19.4 | 23.1 |
| R7 | $C_3H_8+NO_2 \rightarrow C_3H_7\_S+TRANS\_HONO$ | 0 | 31.3 | 14.9 | 19.2 |
| R8 | $C_3H_8+NO_2 \rightarrow C_3H_7\_S+HNO_2$ | 0 | 26.2 | 23.4 | 28.3 |
| R9 | $C_3H_8+NO_2 \rightarrow C_3H_7\_S+CIS\_HONO$ | 0 | 24.7 | 15.8 | 19.6 |
| R10 | $C_4H_{10}+NO_2 \rightarrow C_4H_9\_P+TRANS\_HONO$ | 0 | 34.0 | 19.4 | 23.2 |
| R11 | $C_4H_{10}+NO_2 \rightarrow C_4H_9\_P+HNO_2$ | 0 | 29.8 | 28.0 | 32.3 |
| R12 | $C_4H_{10}+NO_2 \rightarrow C_4H_9\_P+CIS\_HONO$ | 0 | 28.9 | 20.1 | 23.6 |
| R13 | $C_4H_{10}+NO_2 \rightarrow C_4H_9\_T+TRANS\_HONO$ | 0 | 29.6 | 12.6 | 17.9 |
| R14 | $C_4H_{10}+NO_2 \rightarrow C_4H_9\_T+HNO_2$ | 0 | 23.0 | 21.3 | 26.9 |
| R15 | $C_4H_{10}+NO_2 \rightarrow C_4H_9\_T+CIS\_HONO$ | 0 | 22.3 | 13.0 | 18.2 |
| R16 | $C_5H_{12}+NO_2 \rightarrow C_5H_{11}\_S+TRANS\_HONO$ | 0 | 31.1 | 15 | 19.8 |
| R17 | $C_5H_{12}+NO_2 \rightarrow C_5H_{11}\_S+HNO_2$ | 0 | 25.7 | 23.6 | 28.8 |
| R18 | $C_5H_{12}+NO_2 \rightarrow C_5H_{11}\_S+CIS\_HONO$ | 0 | 25.1 | 16.2 | 20.1 |
| R19 | $C_5H_{12}+NO_2 \rightarrow C_5H_{11}\_T+TRANS\_HONO$ | 0 | 29.5 | 12.6 | 18.3 |
| R20 | $C_5H_{12}+NO_2 \rightarrow C_5H_{11}\_T+HNO_2$ | 0 | 22.9 | 21.3 | 27.3 |
| R21 | $C_5H_{12}+NO_2 \rightarrow C_5H_{11}\_T+CIS\_HONO$ | 0 | 21.9 | 13.3 | 18.6 |
| Alkenes + $NO_2$ | | | | | |
| R22 | $C_2H_4+NO_2 \rightarrow C_2H_3\_v(S)+TRANS\_HONO$ | 0 | 38.3 | 28.3 | 31.9 |
| R23 | $C_2H_4+NO_2 \rightarrow C_2H_3\_v(S)+HNO_2$ | 0 | 36.9 | 36.9 | 41.0 |
| R24 | $C_2H_4+NO_2 \rightarrow C_2H_3\_v(S)+CIS\_HONO$ | 0 | 34.3 | 29.3 | 32.3 |
| R25 | $C_3H_6+NO_2 \rightarrow C_3H_5\_\alpha(P)+TRANS\_HONO$ | 0 | 32.2 | 4.9 | 8.4 |
| R26 | $C_3H_6+NO_2 \rightarrow C_3H_5\_\alpha(P)+HNO_2$ | 0 | 24.8 | 13.7 | 17.5 |
| R27 | $C_3H_6+NO_2 \rightarrow C_3H_5\_\alpha(P)+CIS\_HONO$ | 0 | 25.4 | 5.4 | 8.8 |
| R28 | $C_3H_6+NO_2 \rightarrow C_3H_5\_v(S)+TRANS\_HONO$ | 0 | 41.3 | 29.2 | 33.7 |
| R29 | $C_3H_6+NO_2 \rightarrow C_3H_5\_v(S)+HNO_2$ | 0 | 38.3 | 36.3 | 41.8 |
| R30 | $C_3H_6+NO_2 \rightarrow C_3H_5\_v(S)+CIS\_HONO$ | 0 | 36.8 | 29.9 | 34.1 |
| R31 | $C_3H_6+NO_2 \rightarrow C_3H_5\_v(T)+TRANS\_HONO$ | 0 | 35.1 | 24.2 | 29.0 |
| R32 | $C_3H_6+NO_2 \rightarrow C_3H_5\_v(T)+HNO_2$ | 0 | 34.8 | 32.5 | 38.2 |
| R33 | $C_3H_6+NO_2 \rightarrow C_3H_5\_v(T)+CIS\_HONO$ | 0 | 30.9 | 24.9 | 29.3 |
| R34 | $C_4H_8\_2+NO_2 \rightarrow C_4H_7\_\alpha(P)+TRANS\_HONO$ | 0 | 30.1 | 3.8 | 8.1 |
| R35 | $C_4H_8\_2+NO_2 \rightarrow C_4H_7\_\alpha(P)+HNO_2$ | 0 | 23.3 | 12.5 | 17.1 |
| R36 | $C_4H_8\_2+NO_2 \rightarrow C_4H_7\_\alpha(P)+CIS\_HONO$ | 0 | 24.0 | 4.7 | 8.4 |
| R37 | $IC_4H_8+NO_2 \rightarrow IC_4H_7\_\alpha(P)+TRANS\_HONO$ | 0 | 30.8 | 5.7 | 9.9 |
| R38 | $IC_4H_8+NO_2 \rightarrow IC_4H_7\_\alpha(P)+HNO_2$ | 0 | 24.3 | 14.5 | 18.9 |
| R39 | $IC_4H_8+NO_2 \rightarrow IC_4H_7\_\alpha(P)+CIS\_HONO$ | 0 | 24.3 | 6.2 | 10.2 |
| R40 | $C_4H_8\_1+NO_2 \rightarrow C_4H_7\_\alpha(S)+TRANS\_HONO$ | 0 | 29.7 | 1.1 | 5.3 |
| R41 | $C_4H_8\_1+NO_2 \rightarrow C_4H_7\_\alpha(S)+HNO_2$ | 0 | 21.8 | 9.5 | 14.3 |
| R42 | $C_4H_8\_1+NO_2 \rightarrow C_4H_7\_\alpha(S)+CIS\_HONO$ | 0 | 22.3 | 2.3 | 5.6 |
| R43 | $C_5H_{10}+NO_2 \rightarrow C_5H_9\_\alpha(T)+TRANS\_HONO$ | 0 | 28.0 | -1.4 | 3.6 |
| R44 | $C_5H_{10}+NO_2 \rightarrow C_5H_9\_\alpha(T)+HNO_2$ | 0 | 19.4 | 7.4 | 12.6 |
| R45 | $C_5H_{10}+NO_2 \rightarrow C_5H_9\_\alpha(T)+CIS\_HONO$ | 0 | 20.3 | -0.7 | 3.9 |

The reactive energies of H-atom abstraction from alkanes and alkenes are detailed in Table 1, where product complexes are identified for all reactions. For alkanes, the energy required for hydrogen abstraction from identical sites across different reactants remains roughly consistent, including at the primary (P), secondary (S), and tertiary (T) sites. However, the energy barriers significantly differ among these sites, with the order from highest to lowest energy being: P > S > T. This variation is primarily attributed to the influence of the C-C bond interactions. For instance, in the case of the TRANS_HONO product, the relative energy of the TS for $C_3H_8$ at the P site is 34.0 kcal/mol, which is higher than that at the S site (31.3 kcal/mol). In $C_5H_{12}$, the relative energy of TS at the S site (31.1 kcal/mol) is higher than that at the T site (29.5 kcal/mol). Notably, the number of carbon atoms has a minimal impact on the energy barriers at the same carbon site. For alkenes, the energy barrier for H-atom abstraction from the α site is substantially lower than those from the ν site. For example, in the case of the TRANS_HONO product involving $C_3H_6$, the H abstraction reaction from the α(P), ν(S), and ν(T) sites shows the following order of reactive energies, from highest to lowest: ν(S) (41.3 kcal/mol) > ν(T) (35.1 kcal/mol) > α(P) (32.2 kcal/mol).

### 3.3. Rate constant results

The rate coefficients for H-atom abstraction by $NO_2$ species from alkanes and alkenes, calculated across a temperature range of 298.15K to 2000K (presented in the Supplementary Material as Figs. S3-S4), with the fitted Arrhenius rate parameters are presented in Table 2. Notably, except for the reactions involving $C_2H_4$ to form $C_2H_3\_v(S)$ and $C_3H_6$ to form $C_3H_5\_v(T)$, those producing the TRANS_HONO product consistently exhibit the lowest rate coefficients compared to those forming other products. This trend is attributed to the higher energy barriers associated with the TRANS_HONO reactions, as depicted in Figs. S4 and S5. Further, a detailed comparison of the rate coefficients for H-atom abstraction by the $NO_2$ radical from various sites on $C_3H_8$, $C_4H_{10}$, $C_5H_{12}$, and $C_3H_6$ leading to the formation of $HNO_2$ isomers is available in Figs. S6-S9.

Table 2 The rate coefficient for H-atom abstraction by $NO_2$ from alkanes and alkenes to form the respective products and $HNO_2$ isomers (TRANS_HONO, $HNO_2$, CIS_HONO).

| No. | Reaction | A (cm$^3$/mol*s) | n | Ea (cal/mol) |
|---|---|---|---|---|
| Alkanes + $NO_2$ | | | | |
| R1 | $C_2H_6+NO_2 \rightarrow C_2H_5\_P+TRANS\_HONO$ | 8.381E+00 | 3.771 | 31656.03 |
| R2 | $C_2H_6+NO_2 \rightarrow C_2H_5\_P+HNO_2$ | 1.796E+02 | 3.319 | 27454.65 |
| R3 | $C_2H_6+NO_2 \rightarrow C_2H_5\_P+CIS\_HONO$ | 1.000E+00 | 4.001 | 24215.77 |
| R4 | $C_3H_8+NO_2 \rightarrow C_3H_7\_P+TRANS\_HONO$ | 5.463E+01 | 3.524 | 32395.00 |
| R5 | $C_3H_8+NO_2 \rightarrow C_3H_7\_P+HNO_2$ | 6.324E+02 | 3.148 | 28280.96 |
| R6 | $C_3H_8+NO_2 \rightarrow C_3H_7\_P+CIS\_HONO$ | 1.000E+00 | 3.966 | 24385.10 |
| R7 | $C_3H_8+NO_2 \rightarrow C_3H_7\_S+TRANS\_HONO$ | 2.696E+01 | 3.435 | 29724.89 |
| R8 | $C_3H_8+NO_2 \rightarrow C_3H_7\_S+HNO_2$ | 6.947E+02 | 2.969 | 24329.93 |
| R9 | $C_3H_8+NO_2 \rightarrow C_3H_7\_S+CIS\_HONO$ | 1.000E+00 | 3.898 | 21365.91 |
| R10 | $C_4H_{10}+NO_2 \rightarrow C_4H_9\_P+TRANS\_HONO$ | 2.136E+04 | 2.830 | 33310.08 |
| R11 | $C_4H_{10}+NO_2 \rightarrow C_4H_9\_P+HNO_2$ | 2.503E+05 | 2.460 | 29193.57 |
| R12 | $C_4H_{10}+NO_2 \rightarrow C_4H_9\_P+CIS\_HONO$ | 3.997E+00 | 4.052 | 25532.44 |
| R13 | $C_4H_{10}+NO_2 \rightarrow C_4H_9\_T+TRANS\_HONO$ | 1.002E+00 | 3.760 | 27250.29 |
| R14 | $C_4H_{10}+NO_2 \rightarrow C_4H_9\_T+HNO_2$ | 1.002E+00 | 3.819 | 19994.76 |
| R15 | $C_4H_{10}+NO_2 \rightarrow C_4H_9\_T+CIS\_HONO$ | 1.002E+00 | 3.731 | 19265.09 |
| R16 | $C_5H_{12}+NO_2 \rightarrow C_5H_{11}\_S+TRANS\_HONO$ | 1.353E+06 | 1.777 | 31318.68 |
| R17 | $C_5H_{12}+NO_2 \rightarrow C_5H_{11}\_S+HNO_2$ | 1.002E+00 | 3.628 | 22630.73 |
| R18 | $C_5H_{12}+NO_2 \rightarrow C_5H_{11}\_S+CIS\_HONO$ | 8.607E+03 | 2.610 | 23134.42 |
| R19 | $C_5H_{12}+NO_2 \rightarrow C_5H_{11}\_T+TRANS\_HONO$ | 1.002E+00 | 3.632 | 26981.62 |
| R20 | $C_5H_{12}+NO_2 \rightarrow C_5H_{11}\_T+HNO_2$ | 1.002E+00 | 3.705 | 19764.80 |
| R21 | $C_5H_{12}+NO_2 \rightarrow C_5H_{11}\_T+CIS\_HONO$ | 1.002E+00 | 3.659 | 18896.64 |
| | Alkenes + $NO_2$ | | | |
| R22 | $C_2H_4+NO_2 \rightarrow C_2H_3\_v(S)+TRANS\_HONO$ | 1.000E+00 | 3.975 | 35023.47 |
| R23 | $C_2H_4+NO_2 \rightarrow C_2H_3\_v(S)+HNO_2$ | 3.659E+02 | 3.136 | 35734.37 |
| R24 | $C_2H_4+NO_2 \rightarrow C_2H_3\_v(S)+CIS\_HONO$ | 1.000E+00 | 4.017 | 30711.63 |
| R25 | $C_3H_6+NO_2 \rightarrow C_3H_5\_\alpha(P)+TRANS\_HONO$ | 1.000E+00 | 3.582 | 27029.53 |
| R26 | $C_3H_6+NO_2 \rightarrow C_3H_5\_\alpha(P)+HNO_2$ | 1.000E+00 | 3.593 | 20547.99 |
| R27 | $C_3H_6+NO_2 \rightarrow C_3H_5\_\alpha(P)+CIS\_HONO$ | 1.000E+00 | 3.710 | 20783.04 |
| R28 | $C_3H_6+NO_2 \rightarrow C_3H_5\_v(S)+TRANS\_HONO$ | 1.000E+00 | 3.820 | 38545.24 |
| R29 | $C_3H_6+NO_2 \rightarrow C_3H_5\_v(S)+HNO_2$ | 1.000E+00 | 3.698 | 36203.34 |
| R30 | $C_3H_6+NO_2 \rightarrow C_3H_5\_v(S)+CIS\_HONO$ | 1.000E+00 | 3.834 | 33531.74 |
| R31 | $C_3H_6+NO_2 \rightarrow C_3H_5\_v(T)+TRANS\_HONO$ | 1.000E+00 | 3.844 | 32793.44 |
| R32 | $C_3H_6+NO_2 \rightarrow C_3H_5\_v(T)+HNO_2$ | 3.233E+02 | 3.024 | 33717.34 |
| R33 | $C_3H_6+NO_2 \rightarrow C_3H_5\_v(T)+CIS\_HONO$ | 1.000E+00 | 3.771 | 27725.43 |
| R34 | $C_4H_8\_2+NO_2 \rightarrow C_4H_7\_\alpha(P)+TRANS\_HONO$ | 1.000E+00 | 3.490 | 25757.47 |
| R35 | $C_4H_8\_2+NO_2 \rightarrow C_4H_7\_\alpha(P)+HNO_2$ | 1.000E+00 | 3.483 | 19278.77 |
| R36 | $C_4H_8\_2+NO_2 \rightarrow C_4H_7\_\alpha(P)+CIS\_HONO$ | 1.000E+00 | 3.710 | 19729.98 |
| R37 | $IC_4H_8+NO_2 \rightarrow IC_4H_7\_\alpha(P)+TRANS\_HONO$ | 1.000E+00 | 3.738 | 25759.37 |
| R38 | $IC_4H_8+NO_2 \rightarrow IC_4H_7\_\alpha(P)+HNO_2$ | 1.000E+00 | 3.765 | 20181.60 |
| R39 | $IC_4H_8+NO_2 \rightarrow IC_4H_7\_\alpha(P)+CIS\_HONO$ | 1.000E+00 | 3.840 | 20190.04 |
| R40 | $C_4H_8\_1+NO_2 \rightarrow C_4H_7\_\alpha(S)+TRANS\_HONO$ | 1.000E+00 | 3.683 | 25988.52 |
| R41 | $C_4H_8\_1+NO_2 \rightarrow C_4H_7\_\alpha(S)+HNO_2$ | 1.000E+00 | 3.624 | 17495.85 |
| R42 | $C_4H_8\_1+NO_2 \rightarrow C_4H_7\_\alpha(S)+CIS\_HONO$ | 1.000E+00 | 3.665 | 18849.79 |
| R43 | $C_5H_{10}+NO_2 \rightarrow C_5H_9\_\alpha(T)+TRANS\_HONO$ | 1.000E+00 | 3.612 | 25090.44 |
| R44 | $C_5H_{10}+NO_2 \rightarrow C_5H_9\_\alpha(T)+HNO_2$ | 1.000E+00 | 3.642 | 15310.23 |
| R45 | $C_5H_{10}+NO_2 \rightarrow C_5H_9\_\alpha(T)+CIS\_HONO$ | 1.000E+00 | 3.680 | 17078.69 |

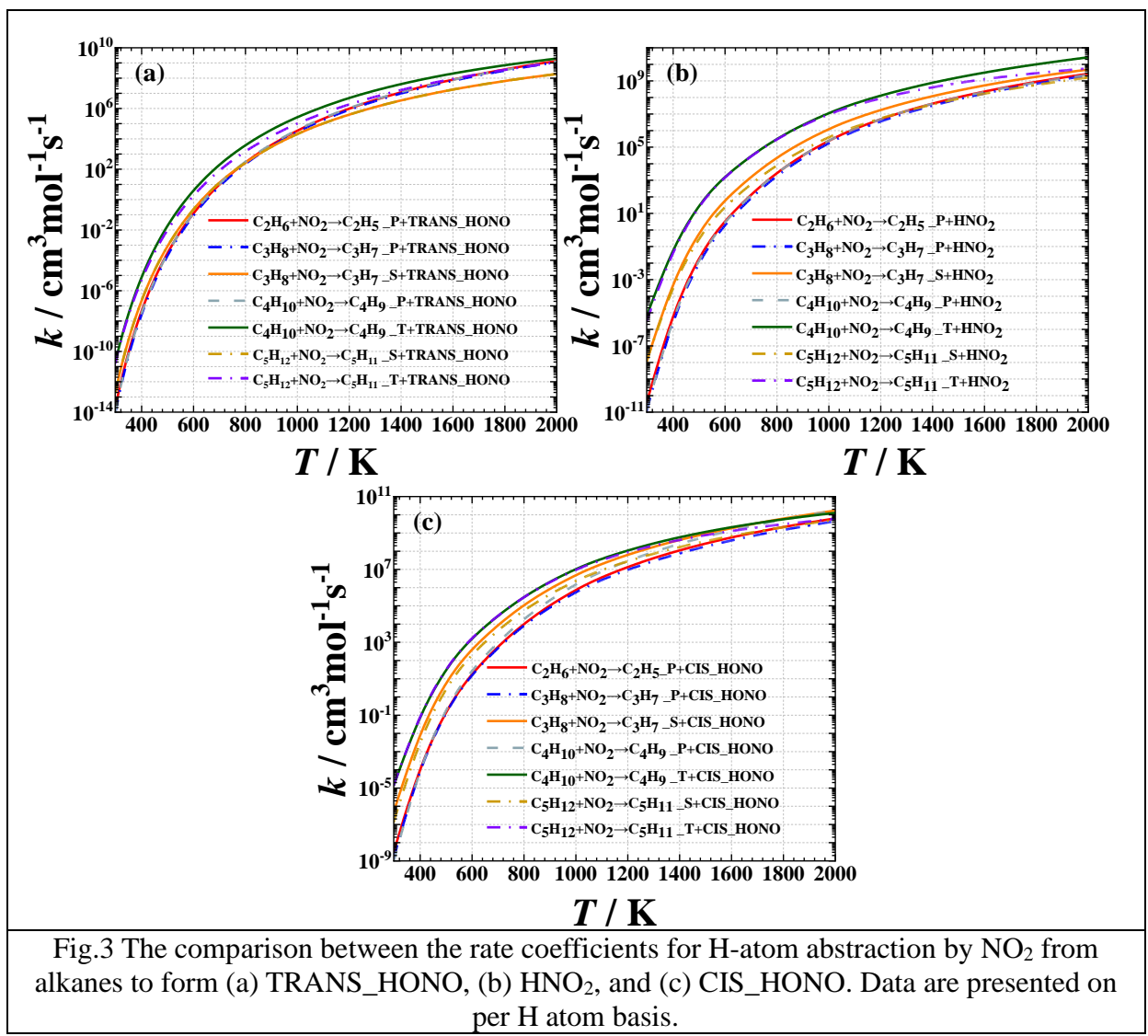

Fig.3 The comparison between the rate coefficients for H-atom abstraction by $NO_2$ from alkanes to form (a) TRANS_HONO, (b) $HNO_2$, and (c) CIS_HONO. Data are presented on per H atom basis.

Figure 3 demonstrates that when H-atom abstraction occurs at the P, S, and T site in different alkanes, the differences in rate coefficients at the same site are within one order of magnitude. At low temperatures, the rate coefficients at the tertiary (T) site are the highest, followed by the secondary (S) site, which is 2-3 orders of magnitude lower than the T site. The coefficients at the primary (P) site are the lowest, about 1-3 orders of magnitude lower than the S site. This distribution is consistent with the energy barriers at the P, S, and T sites. As temperature increases, the disparities in rate coefficients among different sites diminish, eventually narrowing to 1-2 orders of magnitude at 2000 K.

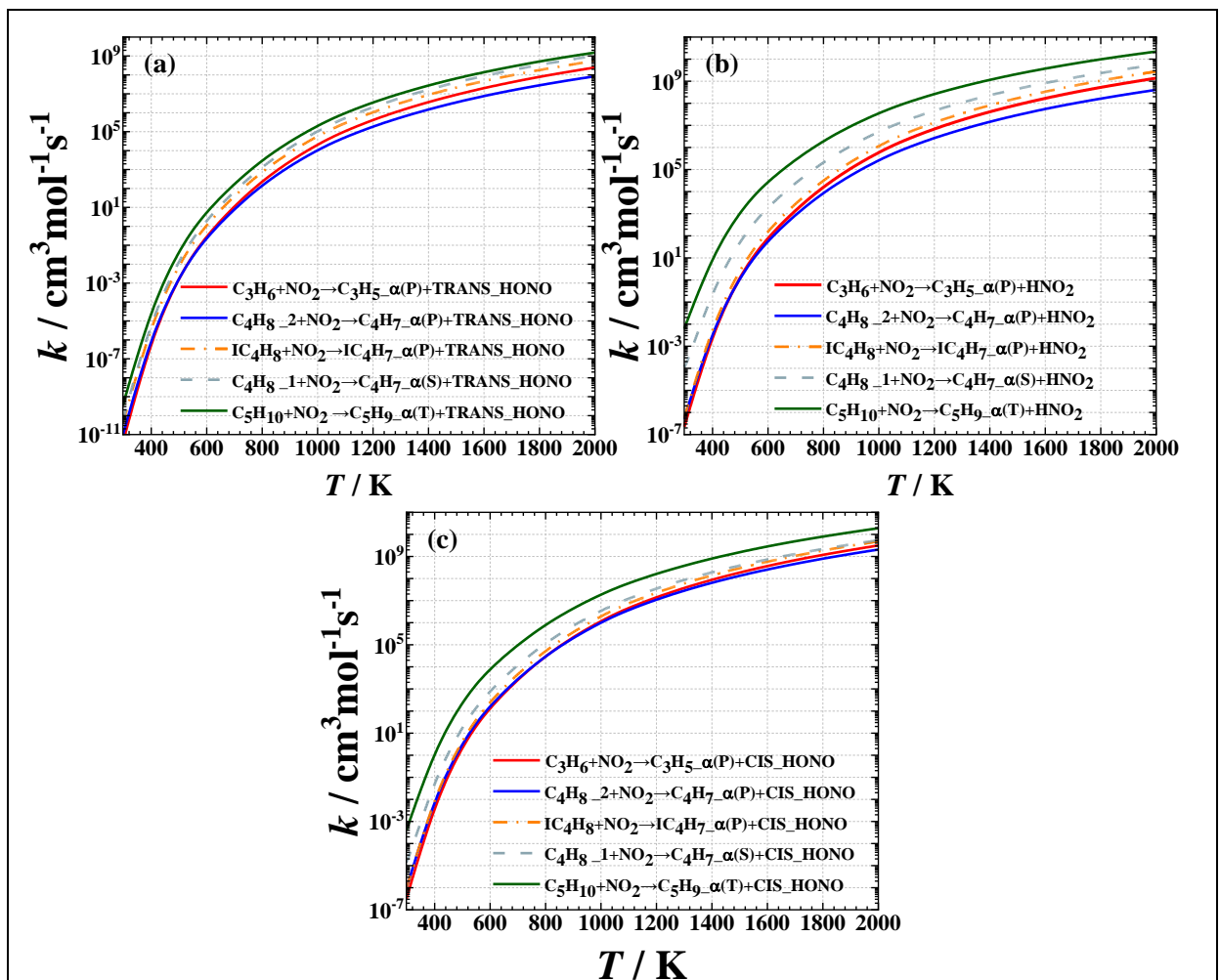

Fig.4 The comparison between the rate coefficients for H-atom abstraction by $NO_2$ at α site from alkenes to form (a) TRANS_HONO, (b) $HNO_2$, and (c) CIS_HONO. Data are presented on per H atom basis.

Figure 4 assesses the rate coefficients at the α site for the P, S, and T sites of alkenes. When the product is TRANS_HONO, namely Fig. 4(a), the difference between rate coefficients at the α(P), α(S), and α(T) sites are relatively small at 298K, within 2 orders of magnitude. The rate coefficients of different sites can be ranked as α(T) > α(S) > α(P), with the difference becoming more pronounced when temperature increases. For the products of $HNO_2$ and CIS_HONO, as depicted in Fig. 4(b) and (c), respectively, the rate coefficient at the P site is 3-5 orders lower than those at S and T sites for $HNO_2$ and 1-3 orders lower than those at S and T sites for CIS_HONO at 298K.

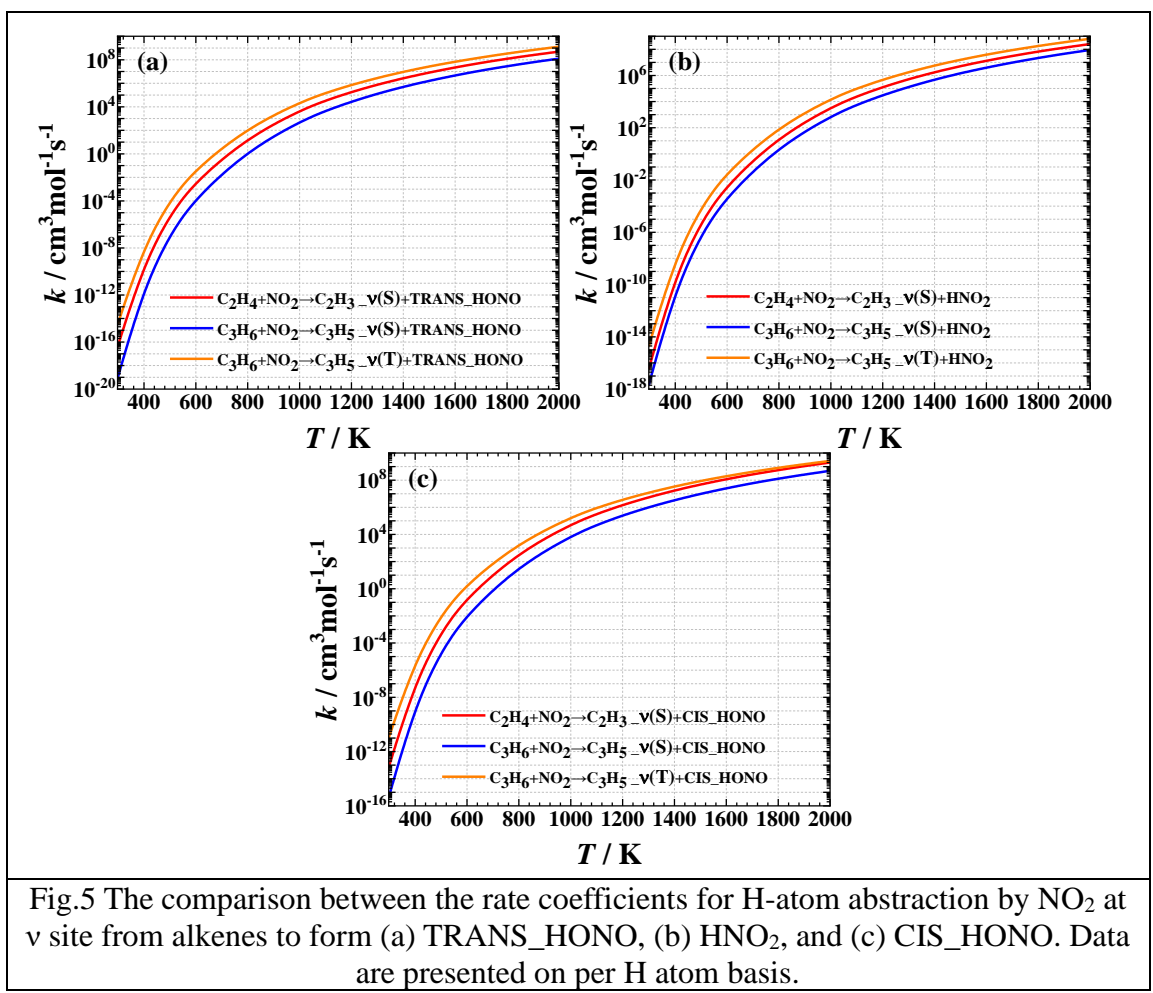

Fig.5 The comparison between the rate coefficients for H-atom abstraction by $NO_2$ at ν site from alkenes to form (a) TRANS_HONO, (b) $HNO_2$, and (c) CIS_HONO. Data are presented on per H atom basis.

Figure 5 compares the rate coefficients for H-atom abstraction by $NO_2$ at the ν site from alkenes, the rate coefficients from fastest to slowest follows: ν(T) of $C_3H_6$ > ν(S) of $C_2H_4$ > ν(S) of $C_3H_6$. Furthermore, the rate coefficients at the ν site are significantly lower than those at the α site, reflecting the strong influences of the C=C bond. Unlike the results presented in Figs. 4 and 5, the rates at the same site (i.e., ν(S)) becomes quite different. Specifically, the rates at the ν(S) site of $C_3H_6$ are 1-2 orders lower than those of $C_2H_4$, regardless of the products formed.

3.4. Branching ratio analysis

Figure 6 illustrates the branching ratios for H-atom abstractions by $NO_2$ from different alkanes. The trends of branching ratios for primary (P) sites (i.e., Figs. 6(a), 6(b) and 6(c)) are similar, where CIS_HONO production occupies nearly 99% of the total abstractions at 298K with minimal contributions from $HNO_2$ and TRANS_HONO production pathways. As temperature increases, the branching ratio of CIS_HONO gradually decreases to 60%-83% at 2000K, while the ratios for $HNO_2$ and TRANS_HONO rise to 10-25% and approximately 6%, respectively. For the S sites, as depicted in Figs. 6(d) and 6(e), the branching ratios are similar to those of the P site, though there are slight qualitative differences for the S site of $C_5H_{12}$. Regarding the T site, as shown in Figs. 6(f) and 6(g), it is evident that the trends of branching ratios differ greatly from those for the P and S sites. At this site, the branching ratios of CIS_HONO and $HNO_2$ intersect at a specific temperature, while that of TRANS_HONO remains low.

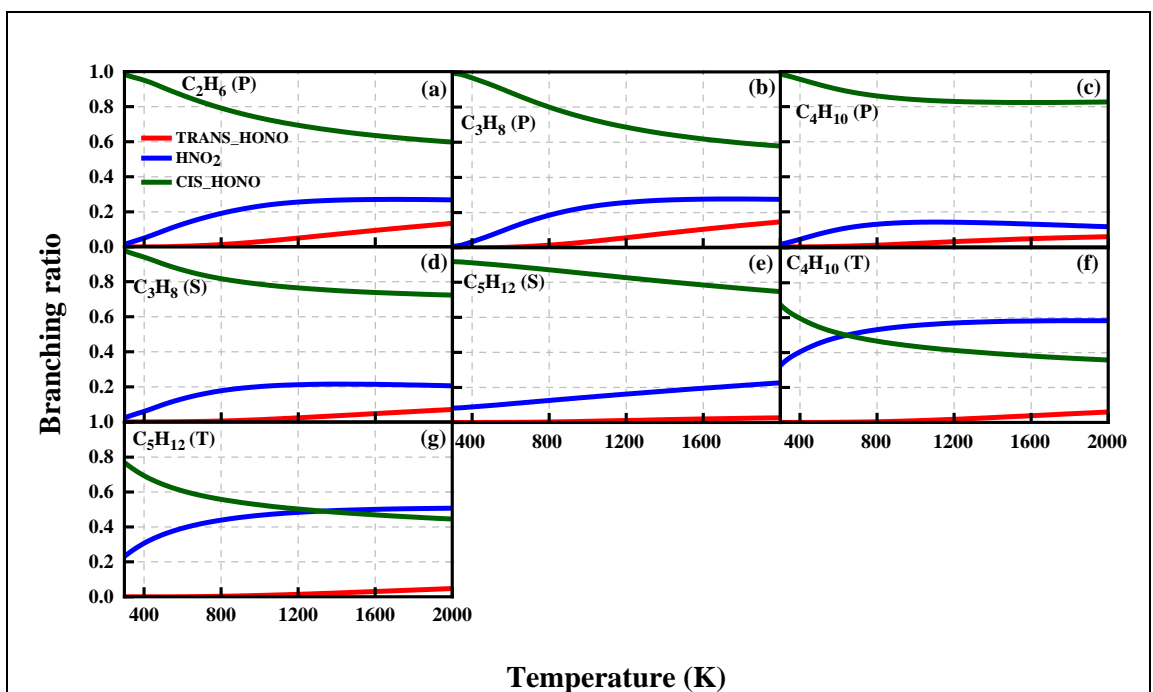

Fig.6 The comparison between the branching ratio for H-atom abstraction by $NO_2$ from alkanes to form $HNO_2$ isomers (TRANS_HONO, $HNO_2$, CIS_HONO), (a) $C_2H_6$(P), (b) $C_3H_8$(P), (c) $C_4H_{10}$(P), (d) $C_3H_8$(S), (e) $C_5H_{12}$(S), (f) $C_4H_{10}$(T), (g) $C_5H_{12}$(T).

Fig.7 assesses the branching ratios for H-atom abstraction by $NO_2$ from alkenes. As can be seen from Figs. 8(a), 8(b) and 8(c), the branching ratio of CIS_HONO for α-P site is slightly higher than that of HNO$_2$. The branching ratio of TRANS_HONO remains the lowest. With rising temperatures, the branching ratios of CIS_HONO and TRANS_HONO increase, while that of HNO$_2$ decreases. The trends become quite different for the α-P and α-S sites, where the branching ratio is dominated by HNO$_2$. The branching ratios of HNO$_2$ and CIS_HONO merge at high temperatures. At the ν-S and ν-T sites (Figs. 8(f), 8(g) and 8(h)), there is, again, a clear shift in branching ratios, where CIS_HONO consistently exhibits the highest branching ratio. The branching ratios of HNO$_2$ and TRANS_HONO is significantly lower and higher than that at the α site throughout the temperature range studied.

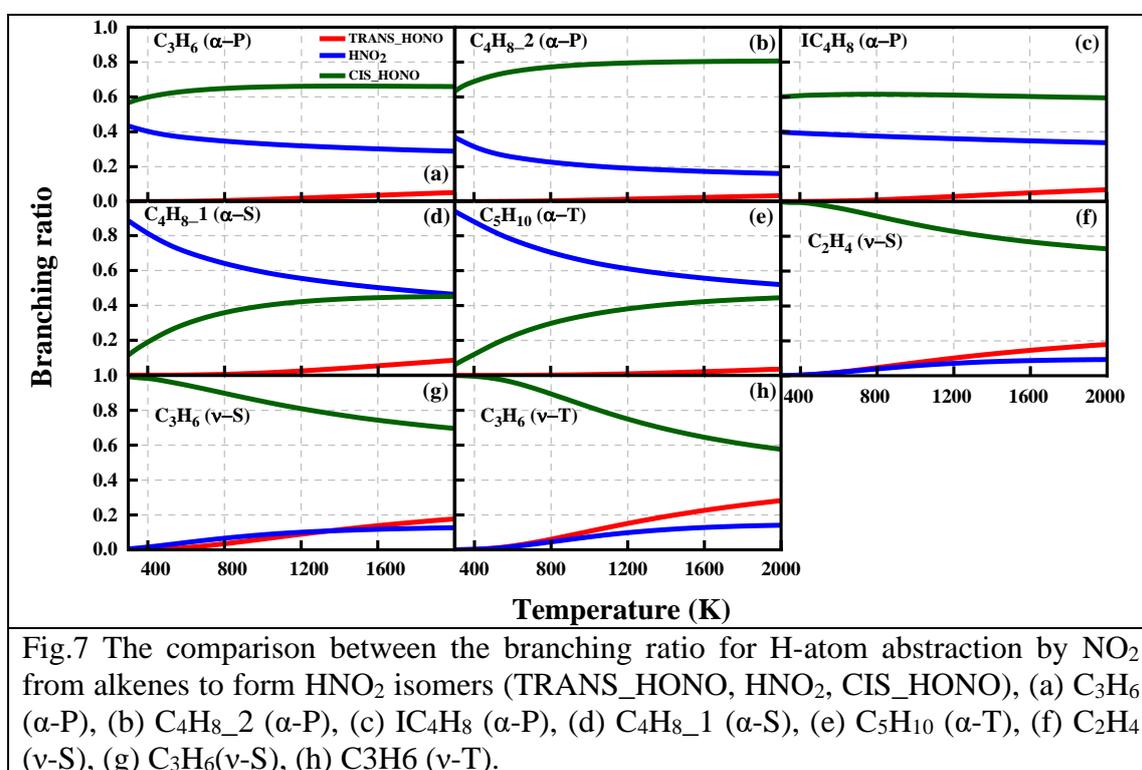

Fig.7 The comparison between the branching ratio for H-atom abstraction by NO$_2$ from alkenes to form HNO$_2$ isomers (TRANS_HONO, HNO$_2$, CIS_HONO), (a) C$_3$H$_6$ (α-P), (b) C$_4$H$_8$_2 (α-P), (c) IC$_4$H$_8$ (α-P), (d) C$_4$H$_8$_1 (α-S), (e) C$_5$H$_{10}$ (α-T), (f) C$_2$H$_4$ (ν-S), (g) C$_3$H$_6$(ν-S), (h) C3H6 (ν-T).

3.5. Rate rules

The consistent trends at different sites on the same molecule and at the same site on different molecules offer an opportunity to derive the rate rules for H-atom abstraction reactions by NO$_2$ from alkanes and alkenes. In this regard, the branching ratios at different sites of different molecules are presented as functions of carbon number at four different temperatures (i.e., 600K, 800K, 1000K, 1200K), which are shown in Fig. 8. It is clear from Fig. 8 that the branching ratios, when plotted against carbon numbers, follow a linear trend, with distinctive differences between different carbon sites. For instance, the branching ratios with respect to carbon number for CIS_HONO follow Y=0.0166X+0.8323, Y=0.0129X+0.828 and Y=0.0956X+0.1288 for the H-atom abstractions from the P, S and T sites of alkanes, respectively. The branching ratio for TRANS_HONO can be correlated with a single linear fit, regardless of the abstraction sites. These trends are consistent at all temperatures. Nevertheless, as temperature increases, the fitted lines for the P and S start to deviate from each other, which is observed for both CIS_HONO and $HNO_2$.

Figure 9 illustrates the branching ratios as functions of carbon numbers for different alkenes, along with the respective linear fitting lines. Similar to those observed in Fig. 8, there are clear site-to-site differences in trend lines for $HNO_2$ and CIS-HONO, whereas the branching ratios of TRANS_HONO can be described by a single linear fit for all abstractions sites. The excellent agreement between the fitting lines and the computed branching ratios, as shown in Figs. 8 and 9, clearly warrants the use of these fitting lines as rate rules for defining the branching ratios for heavier alkanes and alkenes.

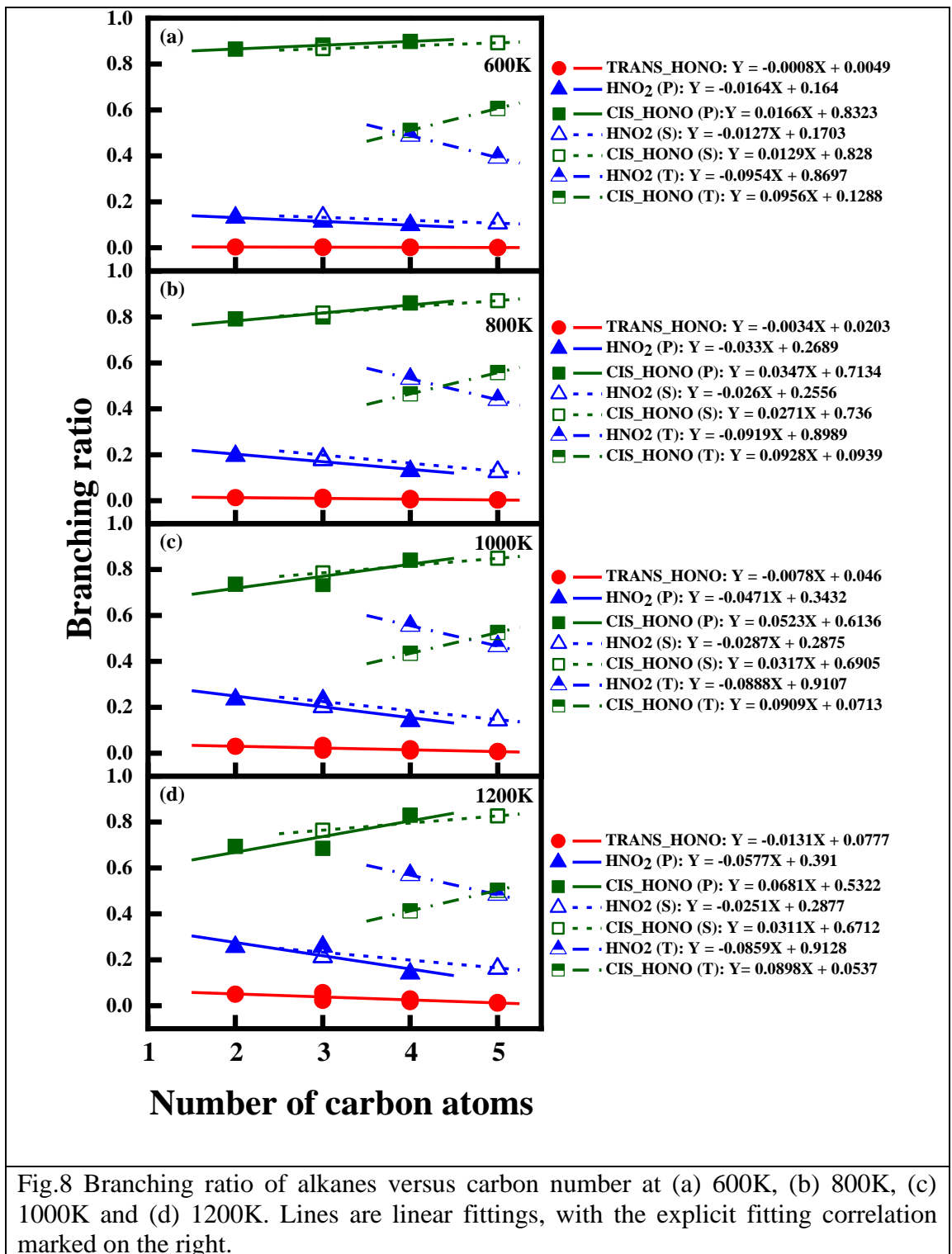

Fig.8 Branching ratio of alkanes versus carbon number at (a) 600K, (b) 800K, (c) 1000K and (d) 1200K. Lines are linear fittings, with the explicit fitting correlation marked on the right.

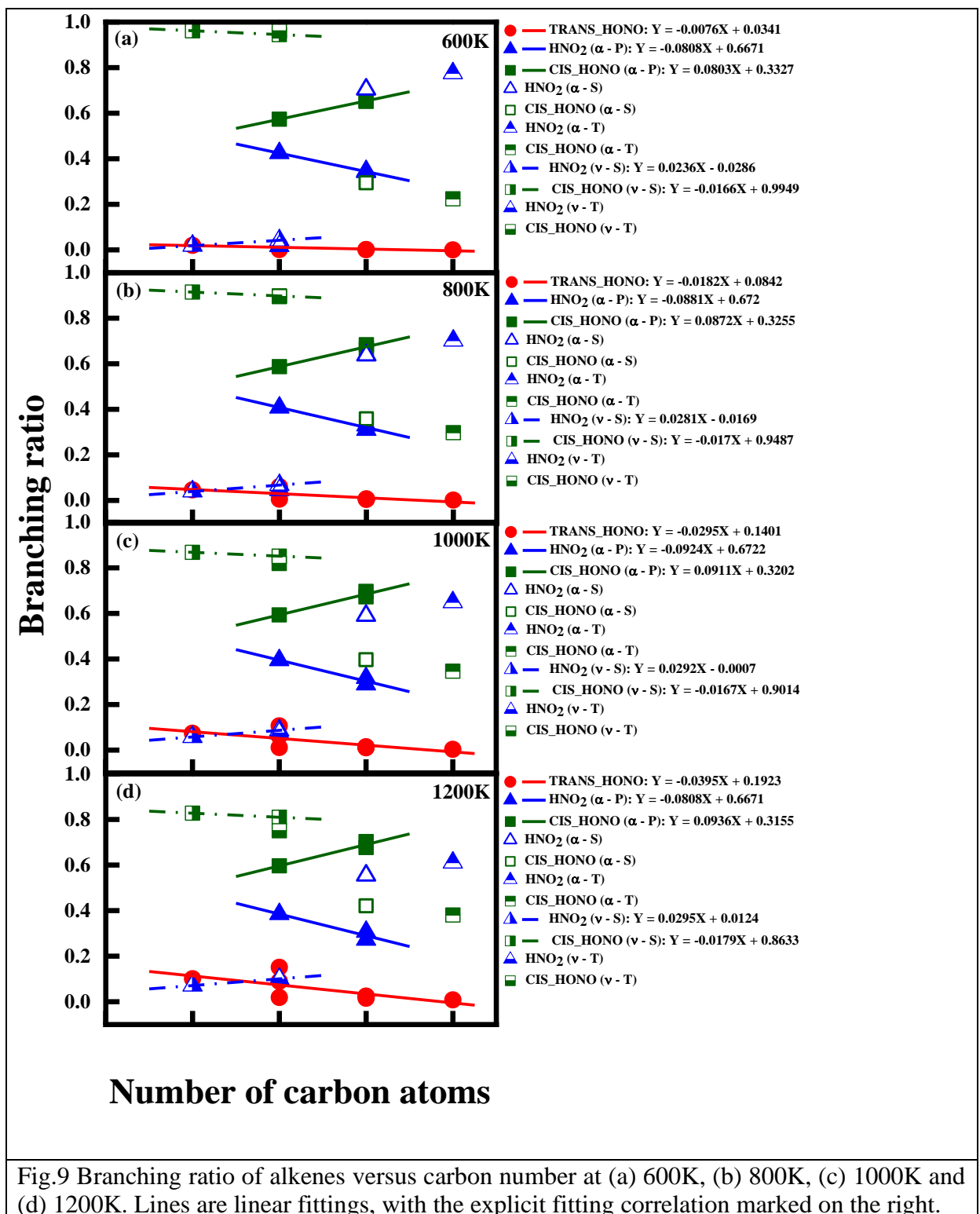

Fig.9 Branching ratio of alkenes versus carbon number at (a) 600K, (b) 800K, (c) 1000K and (d) 1200K. Lines are linear fittings, with the explicit fitting correlation marked on the right.

3.6 Model implementation and implications

To demonstrate the influence of newly calculated rate constants on model performance, the ignition delay time (IDT) of the investigated species at a pressure of 40 bar, equivalence ratio of 1, temperature range of 600-1200 K, and with 1000 ppm $NO_2$ doped is calculated based on the kinetic model of Cheng et al. [31]. This condition mimics the condition that has been used to study NOx blending effects in the tpRCM at Argonne National Laboratory [32,33]. The simulation results calculated using the original model are marked as 'original', while the results calculated using the model with newly calculated rate constants are marked as 'updated', which are compared and analyzed as below.

As shown in Fig. 10, the influence of H-atom abstraction by $NO_2$ reactions on the IDT of $C_3H_8$, $C_3H_6$, and $IC_4H_8$ is obvious. It is worth noting that H-atom abstractions by $NO_2$ reactions already exist in the original model for $C_3H_6$ and $IC_4H_8$, but without CIS_HONO and TRANS_HONO declared. These rates were analogized from the H-atom abstraction by $NO_2$ from $C_2H_4$ [34] which was originally estimated based on the H-atom abstraction by $NO_2$ from $CH_4$ [35]. In addition, the same rate parameters were assigned to different abstraction sites. Based on the results presented in Figs. 6–9 where considerably different rate parameters are obtained between alkanes and alkenes and between different sites, it is obvious that the analogy in the original kinetic model was not accurate. These previous rate constants are replaced by the rate constants calculated by this study. On the other hand, H-atom abstractions by $NO_2$ from $C_3H_8$ are absent from the original model. As such, the calculations from this study are directly added to the original kinetic model. As can be seen from Fig. 10(a), the addition of H-atom abstraction reactions by $NO_2$ promotes the autoignition reactivity of $C_3H_8$, leading to greatly shortened IDT, especially at lower temperatures. There is also a strong promoting effect on the NTC (negative temperature coefficient) behavior. For $C_3H_6$ and $IC_4H_8$ (Fig. 10(b) and 10(c), respectively), incorporating the new reactions greatly inhibits reactivity, leading to longer IDT, while seemingly inhibiting the NTC behavior. Quantitatively speaking, the influences induced by these reactions are significant, which can lead to changes in IDT by up to two magnitude orders.

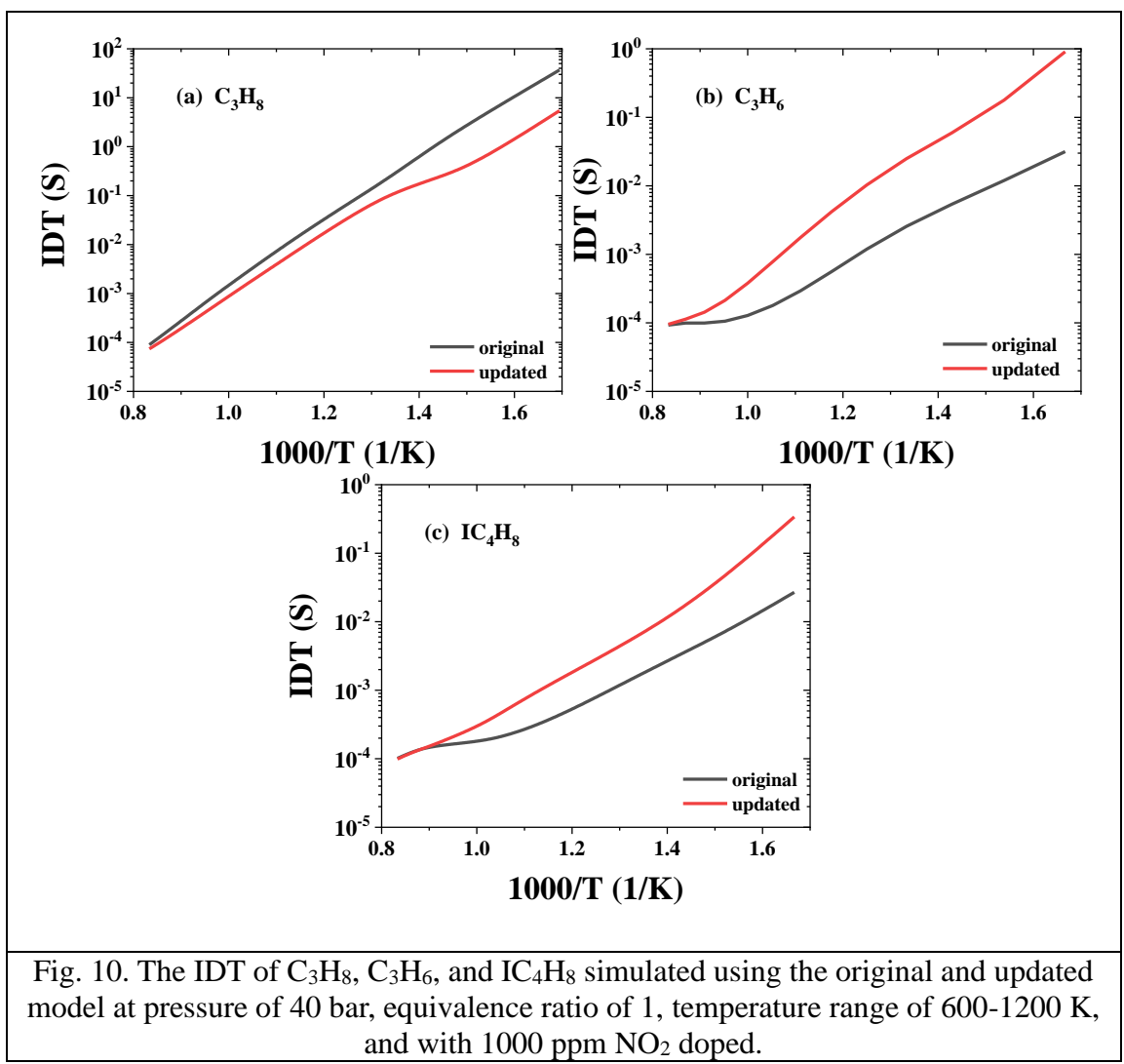

Fig. 10. The IDT of $C_3H_8$, $C_3H_6$, and $IC_4H_8$ simulated using the original and updated model at pressure of 40 bar, equivalence ratio of 1, temperature range of 600-1200 K, and with 1000 ppm $NO_2$ doped.

The impact of H-atom abstractions by $NO_2$ from other molecules on model predictions has also been investigated. The rate parameters calculated from this study are incorporated into the original model [31] individually for the remaining studied species, and the original and updated models are further used to conduct autoignition modeling in ZeroRK [30] at the same conditions as those used in Fig. 10. The relative change in simulated IDT, computed as $|IDT_{updated} - IDT_{original}|/IDT_{original}$, are computed and used to indicate the level of change in model predictions with the updated rate parameters. The results are summarized in Fig. 11, with seven trajectories shown for the seven remaining species. It is clear that the impact of the studied reactions is profound, with the change in simulated IDT ranging from 5% to 45%.

The largest change in simulated IDT is observed for $C_2H_6$ in alkanes and for $C_5H_{10}$ in alkenes. The change in model reactivity exhibits a non-monotonic dependence on temperature, with the highest change observed in the NTC region, indicating a higher impact of $NO_x$ interaction chemistry within this temperature range. This agrees with the experiments reported by Cheng et al. [32], where the measured IDT at the similar conditions (e.g., 40 bar and equivalence ratio of 1.0) also displayed the highest impact from $NO_x$ doping at 750 – 900K for alkenes.

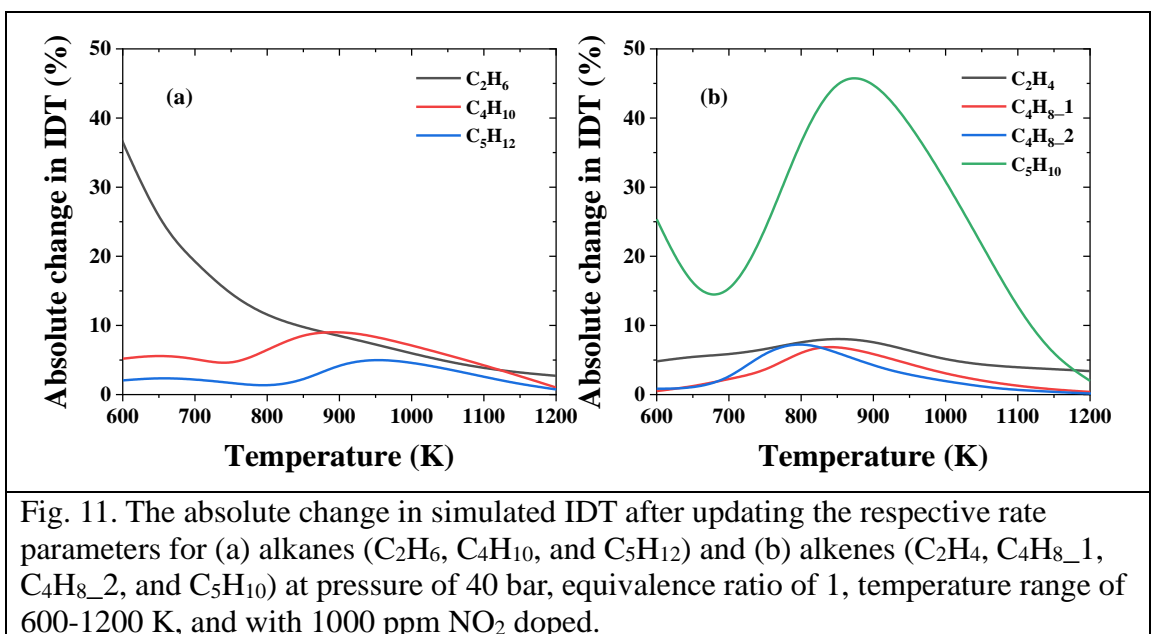

Fig. 11. The absolute change in simulated IDT after updating the respective rate parameters for (a) alkanes ($C_2H_6$, $C_4H_{10}$, and $C_5H_{12}$) and (b) alkenes ($C_2H_4$, $C_4H_8\_1$, $C_4H_8\_2$, and $C_5H_{10}$) at pressure of 40 bar, equivalence ratio of 1, temperature range of 600-1200 K, and with 1000 ppm $NO_2$ doped.

3.7 Sensitivity and flux analysis

To reveal the underlying kinetics governing the model reactivity after incorporating the updated rate parameters, sensitivity analysis is further conducted at the same conditions as in Fig. 10 and 11, but only at two representative temperatures, i.e., 700 and 1100 K. Two representative species, namely $C_3H_8$ and $C_3H_6$, are selected, since these species display the largest change in simulated IDT for alkanes and alkenes, respectively. The sensitivity analysis coefficient is defined as $S_{rel} = \ln(\frac{\tau^\Delta}{\tau})/\ln(\frac{k^\Delta}{k})$, where $\tau^\Delta$ is the main IDT after multiplying the original rate constant by 2, i.e., $k^\Delta = 2 * k$, and $\tau$ is the original ignition delay time. The negative sensitivity coefficient indicates the promotion effect, while the positive sensitivity coefficient indicates the inhibition effect.

Figure 13 shows the sensitivity analysis results for $C_3H_8$ at the investigated conditions. It is obvious from Fig. 13 that the H-atom abstractions by $NO_2$ from $C_3H_8$ greatly promote the ignition reactivity of $C_3H_8$ at both 700 K and 1100 K, as indicated by the negative sensitivity coefficients of $C_3H_8+NO_2=C_3H_7\_S+CIS\_HONO$, $C_3H_8+NO_2=C_3H_7\_P+CIS\_HONO$, and $C_3H_8+NO_2=C_3H_7\_S+HNO_2$. Among these sensitive reactions, $C_3H_8+NO_2=C_3H_7\_S+CIS\_HONO$ has the highest absolute sensitivity coefficient value, which is consistent with the results shown in Fig. 6 where the CIS_HONO pathway has the highest branching ratio. The differences in sensitivity coefficients between the updated and original models diminish from 700 K to 1100 K, indicating a less significant impact of the updated rate parameters on model reactivity at higher temperatures. This is also supported by the results in Fig. 10(a) where the simulated IDT is advanced by less magnitude at higher temperatures.

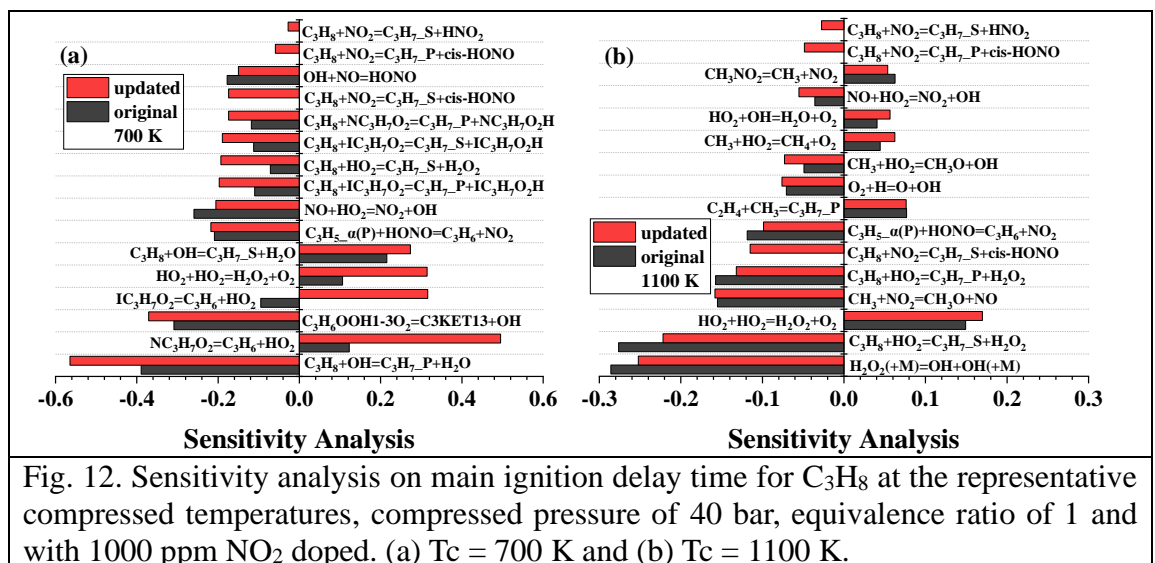

Fig. 12. Sensitivity analysis on main ignition delay time for $C_3H_8$ at the representative compressed temperatures, compressed pressure of 40 bar, equivalence ratio of 1 and with 1000 ppm $NO_2$ doped. (a) Tc = 700 K and (b) Tc = 1100 K.
25

The sensitivity analysis results for $C_3H_6$ are shown in Fig. 13. At Tc = 700 K (Fig. 13(a)), it is found that the promotion effect of $C_3H_6+NO_2=C_3H_5\_\alpha(P)+HONO$ is weakened in the updated model, while the inhibiting effects of the $NO_x$ interaction reactions (e.g., $CH_3NO_2=CH_3+NO_2$, $CH_3O+NO_2=HONO+CH_2O$ and $NO_2+HO_2=HNO_2+O_2$) are enhanced. These changes explain the considerably promoted ignition reactivity as shown in Fig. 10(b). At Tc = 1100 K (Fig. 13(b)), $C_3H_5\_\alpha(P)+NO_2=C_3H_5O+NO$ becomes the second most promotion reaction. The promoting effect is also weakened in the updated model. This weakened promoting effect, along with those from other NOx interacting reactions such as $CH_3O_2+NO=CH_3O+NO_2$, results in the reduced ignition reactivity at this temperature.

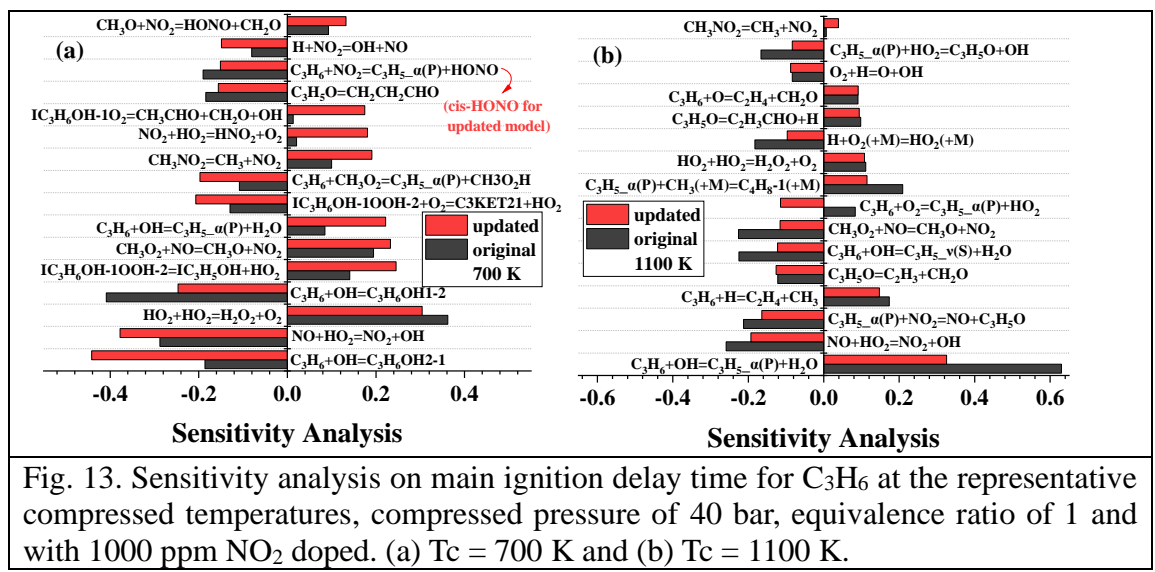

Fig. 13. Sensitivity analysis on main ignition delay time for $C_3H_6$ at the representative compressed temperatures, compressed pressure of 40 bar, equivalence ratio of 1 and with 1000 ppm $NO_2$ doped. (a) Tc = 700 K and (b) Tc = 1100 K.

Flux analysis is further conducted under the same conditions as in Figs. 12 and 13. The results for $C_3H_8$ at 1% fuel consumption are summarized in Fig. 14, while that for $C_3H_6$ is shown in Fig. S10. The numbers shown are in percent, which are computed as the ratio of the rate of consumption (or production) for that pathway to the total rate of consumption (or production). The H-atom abstraction reactions by $NO_2$ are absent from the original model. Thus, the fluxes through these pathways are zero for the original model, as can be seen in Fig. 14. At Tc = 700 K, after the addition of these abstraction reactions, 6.4% and 1.2% of $C_3H_8$ are consumed through the H-atom abstraction by $NO_2$ at the secondary site to form CIS_HONO and $HNO_2$ respectively. Besides, 1.2% of $C_3H_8$ is consumed by $NO_2$ at its primary site to form C$_3$H$_8$_P and CIS_HONO. These pathways are reactivity-promoting reactions, as can be seen from Fig. 12(a), which are expected to increase model reactivity. Another obvious change with the updated rate parameters is the shifted fluxes for the consumption of $CH_2O$, where the flux through $CH_2O+NO_2=HCO+CIS\_HONO$ is reduced by about 11% and 7% at Tc = 700 and 1100 K, respectively. When comparing the results at Tc = 700 K with those at 1100 K, $NO_2$ abstracting H-atom from the second site of $C_3H_8$ decreases to 3.3% and even 0%. Though the primary site abstraction reaction increases slightly from 1.2% to 1.4%, the total dehydrogenation of $C_3H_8$ by $NO_2$ is weakened at 1100 K. Therefore, the influence of $NO_2$ abstracting H-atom reactions on $C_3H_8$ is weakened at higher temperatures, as observed in Fig. 10(a).

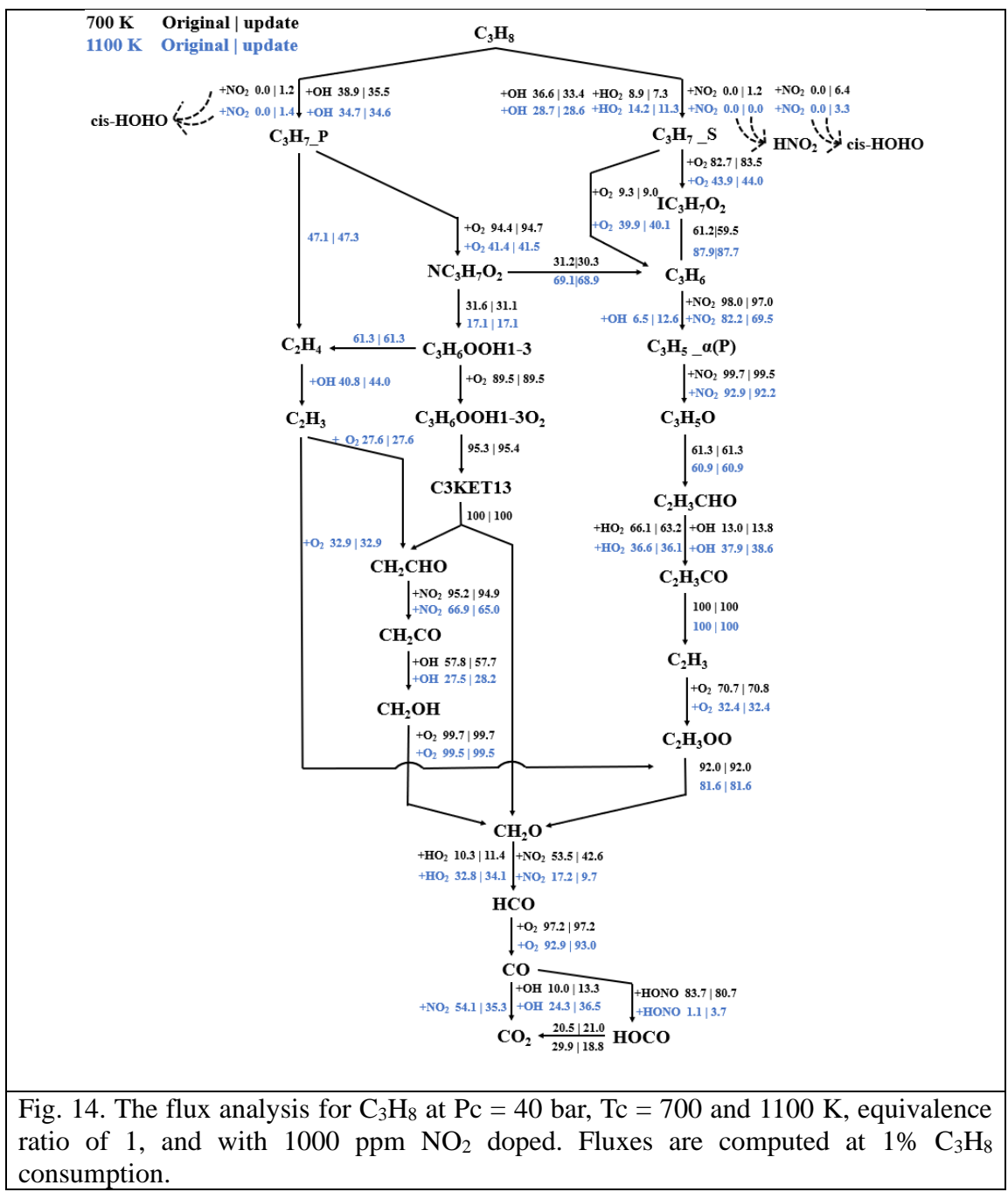

Fig. 14. The flux analysis for $C_3H_8$ at Pc = 40 bar, Tc = 700 and 1100 K, equivalence ratio of 1, and with 1000 ppm $NO_2$ doped. Fluxes are computed at 1% $C_3H_8$ consumption.

## 4 Conclusions

This study presents a comprehensive study on the H-atom abstraction by $NO_2$ from $C_2$-$C_5$ alkanes and alkenes, encompassing 15 hydrocarbons and 3 $HNO_2$ isomers (namely TRANS_HONO, $HNO_2$, and CIS_HONO) with 45 reactions. Employing advanced quantum chemistry computation methods, the bond dissociation energies, potential energy surfaces and rate coefficients are obtained, with the latter subsequently incorporated into a detailed chemistry model for comprehensive chemical kinetic modeling analysis. Branching ratios are further analyzed from the perspective of both carbon site and carbon number. The key findings from this study include:

1. The energy barriers associated with the TRANS_HONO product are consistently higher than those for $HNO_2$ and CIS_HONO across all sites. At the primary (P), secondary (S), and tertiary (T) sites of alkanes and alkenes, the energy barriers decrease in the order: P > S > T. Although the C=C functional group generally lowers the energy barriers for abstracting the adjacent hydrogen atoms at the α carbon site, H-atom abstractions at the C=C sites exhibit significantly higher energy barriers than all other sites.

2. Branching ratios of the pathways forming $HNO_2$, CIS_HONO and TRANS_HONO varies between different species and between different carbon sites on the same molecule, with the CIS_HONO-producing pathway being the most dominant for most species. Nevertheless, further analysis indicates that the branching ratios exhibit clear linear dependences on carbon numbers for the same carbon site. Different rate rules have been proposed for various sites, with can be used to analogize rate parameters to heavier hydrocarbons (e.g., >C5).

3. H-atom abstractions by $NO_2$ are found to greatly affect model reactivity, with incorporating the updated rate parameters generally promoting model reactivity. Sensitivity and flux analysis reveal the important roles of this type of reactions in consuming fuel molecules and the key derivatives such as $CH_2O$, highlighting the necessity of accurately describing the respective kinetics in existing chemistry models.

## Acknowledgments

This material is based on work supported by the Research Grants Council of Hong Kong

Special Administrative Region, China, under PolyU P0046985 for ECS project funded in 2023/24 Exercise and P0050998, and by the Natural Science Foundation of Guangdong Province under 2023A1515010976 and 2024A1515011486.

**Declaration of Competing Interests**

The authors declare no competing interests.